\documentclass[aps,prb,twocolumn,groupedaddress,showpacs]{revtex4}

\usepackage[latin1]{inputenc}   % only if you use accented, non-english, characters
\usepackage{latexsym}
\usepackage{graphicx}
\usepackage{hyperref}

 \providecommand{\dprod}{\! \cdot \!}%

\begin{document}

%\preprint{}

\title{Choice of the best geometry to explain physics}


\author{José B. Almeida}
\email{bda@fisica.uminho.pt}

\thanks{The author wishes to thank Frank Potter, from Sciencegems.com,
for the enlightening discussions and corrections to the text and
equations.}
\affiliation{Universidade do Minho, Physics Department, Campus de
Gualtar, 4710-057 Braga, Portugal}


\date{\today}

\begin{abstract}
Choosing the appropriate geometry in which to express the equations
of fundamental physics can have a determinant effect on the
simplicity of those equations and on the way they are perceived. The
point of departure in this paper is the geometry of 5-dimensional
spacetime, where monogenic functions are studied. Monogenic
functions verify a very simple first order differential equation and
the paper demonstrates how they generate the line interval of
special relativity, as well as the Dirac equation of quantum
mechanics. Monogenic functions act as a unifying principle between
those two areas of physics, which is in itself very significant for
the perception one has of them. Another consequence is the
possibility of studying the same phenomena in Euclidean
4-dimensional space, providing a different point of view to physics,
from which one has an unusual and enriching perspective.
\end{abstract}

%\keywords{}

\maketitle

\section{Introduction}

There is a general consensus among physicists about what is a
physical theory. The point of departure is a set of principles or
axioms which are unproven statements, whose validity is sustained on
the consistency of the whole theory and its ability to make correct
predictions. Using accepted rules of mathematics and logic it is
possible to derive consequences from the set of principles, some of
which are observables and can be confronted with experiment and or
observation of the physical world. In general every theory has its
own application domain, that is, a set of conditions where it is
capable of providing verifiable predictions.

No physical theory has yet been formulated whose application domain
is universal and the search for a unified theory of physics is a
strong motivation for many researchers. The goal is to establish a
reduced number of principles from which one could derive a formalism
applicable to physics of all scales, from particles to the cosmos,
and to all times, from the origin of the Universe, through the
present time, allowing predictions for the Universe's future.

In this paper we will have a brief look at how a carefully chosen
point of departure in geometry can lead us to several important
equations of fundamental physics, using only geometrical reasoning.
We will derive equations with significance both for General
Relativity and Quantum Mechanics, in order to emphasize the
potential unifying power of this approach to physics.

\begin{figure}[htb]
\vspace{11pt} \centerline{\includegraphics[scale=1]{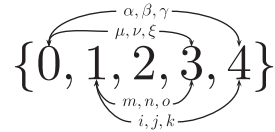}}

\caption{\label{f:indices} Indices in the range $\{0,4\}$ will be
denoted with Greek letters $\alpha, \beta, \gamma.$ Indices in the
range $\{0,3\}$ will also receive Greek letters but chosen from
$\mu, \nu, \xi.$ For indices in the range $\{1,4\}$ we will use
Latin letters $i, j, k$ and finally for indices in the range
$\{1,3\}$ we will use also Latin letters chosen from $m, n, o.$ }
\end{figure}

Before we begin our exposition a  notation issue must be resolved.
We will be dealing with 5-dimensional space but we are also
interested in two of its 4-dimensional subspaces and one
3-dimensional subspace; ideally our choice of indices should clearly
identify their ranges in order to avoid the need to specify the
latter in every equation. The diagram in Fig.\ \ref{f:indices} shows
the index naming convention used in this paper;
Einstein's summation convention will be adopted as well as the
compact notation for partial derivatives $\partial_\alpha =
\partial/\partial x^\alpha.$

\section{Choice of geometry}

Our purpose being to establish geometrical relationships which can
be read as equations of physics, the choice of an appropriate
geometry becomes crucial, no less than an educated assignment of
coordinates to physical entities. Here we will postulate the
geometry as first principle and in doing so we will not try to find
any special justification for that particular choice. The history of
trials leading to the present postulates can be traced through
various papers by the author; the interested reader can find
guidance for such search in the webpage
\url{http://bda.planetaclix.pt}, although our recommendation for
complementary reading is centred in three recent papers
\cite{Almeida04:4, Almeida05:3, Almeida05:1}.

In the introduction we gave a hint that we would be using
5-dimensional space and this is actually true, because we have found
that 5 is the smallest number of dimensions that one needs in order
to find the topology and symmetries that can produce equations
applicable to physics. In the scope of this paper we need 5
dimensions in order to establish a unifying principle from which
both special relativity and quantum mechanics can be derived;
however there are other reasons for this choice, which have to do
with the incorporation of the standard model gauge group symmetries
and a possible hyperspherical symmetry of the
Universe.\cite{Almeida05:1, Almeida05:3} This 5-dimensional space
can be designated as \emph{5-dimensional spacetime} because one of
its dimensions is associated with a frame vector with negative
norm.\endnote{For any space it is possible to define a set of
orthogonal frame vectors in a number equal to the the number of
spatial dimensions; if one of these frame vectors has negative
length, the associated dimension is called a time dimension. The
frame is orthonormed when all its vectors have length $\pm1$.
Alternatively one can also have only one positive length frame
vector, in which case this corresponds to the time dimension.} We
can now add that dimension $0$ will be associated with physical or
Compton time $t$, the time measured by the Compton frequency of
elementary particles, while coordinate $4$ will be associated with
proper or cosmological time $\tau$; these associations will become
clearer further down (see Appendix \ref{units}.) For the moment we
will consider only geometry without any physical implications.

We will need a reference frame associated with the set of
coordinates and we will thus assume an orthonormed frame of vectors,
designated by $\sigma_\alpha$; these vectors are such that
$\sigma_\alpha \dprod \sigma_\beta = 0$, where the dot represents
the inner product and $\alpha \neq \beta$. Furthermore, $\sigma_0$
has norm minus unity, given by $(\sigma_0)^2 = \sigma_0 \dprod
\sigma_0 = -1$ and all the others have unit norm.

An elementary displacement $\mathrm{d}x$ is given by
\begin{equation}
    \label{eq:displacement}
    \mathrm{d} x = \sigma_\alpha \mathrm{d} x^\alpha.
\end{equation}
As usual the square of a vector equals the square of its length,
which we apply to the elementary displacement
\begin{equation}
    \label{eq:null}
    (\mathrm{d} x)^2 = -(\mathrm{d} x^0)^2 + \sum_i (\mathrm{d} x^i)^2.
\end{equation}
Because we have chosen this space to have one negative norm frame
vector, the length of a vector is not necessarily positive and it
can even be zero; we will explore this possibility at great length.

\section{Monogenic functions and waves}

We will now introduce the \emph{vector derivative} defined by
\begin{equation}
    \nabla = \sigma^\alpha \partial_\alpha;
\end{equation}
here we used $\sigma^\alpha$ to represent the \emph{reciprocal
frame} such that $\sigma^\alpha \dprod \sigma_\beta =
\delta^\alpha_\beta$, where $\delta^\alpha_\beta$ is the Kronnecker
delta. One sees easily that $\sigma^0 = -\sigma_0$ and $\sigma^i =
\sigma_i$. It turns out that there is a class of functions of great
importance, called \emph{monogenic functions},\cite{Doran03}
characterized by having null vector derivative; a function $\psi$ is
monogenic if and only if
\begin{equation}
    \label{eq:monogenic}
    \nabla \psi = 0.
\end{equation}
These functions are not usually scalars and we will say a bit more
about them later on but for now let us define the scalar Laplacian
operator $\nabla^2 = \nabla \cdot \nabla$. The Laplacian is just the
sum of second order partial derivatives with respect to all the
coordinates, the term corresponding to coordinate $0$ having a
negative sign $\nabla^2 = -\partial_{00} + \sum
\partial_{ii}$.

A monogenic function has by necessity null Laplacian, as can be seen
by dotting Eq.\ (\ref{eq:monogenic}) with $\nabla$ on the left. We
are then allowed to write
\begin{equation}
    \label{eq:solution}
    \sum_i \partial_{ii} \psi = \partial_{00} \psi.
\end{equation}
This can be recognized as a wave equation in the 4-dimensional space
spanned by $\sigma_i$ which will accept plane wave type solutions of
the general form
\begin{equation}
    \label{eq:psidef}
    \psi = \psi_0 \mathrm{e}^{\mathrm{i} (p_\alpha x^\alpha + \delta)},
\end{equation}
where $\psi_0$ is an amplitude whose characteristics we shall not
discuss for now, $\delta$ is a phase angle and $p_\alpha$ are
constants such that
\begin{equation}
    \label{eq:pnull}
    \sum_i (p_i)^2 - (p_0)^2 = 0.
\end{equation}

By setting the argument of $\psi$ constant in Eq.\ (\ref{eq:psidef})
and differentiating we can get the differential equation
\begin{equation}
    \label{eq:nullcond}
    p_\alpha \mathrm{d}x^\alpha =  0.
\end{equation}
The first member can equivalently be written as the inner product of
the two vectors $p \dprod \mathrm{d}x = 0$, where $p = \sigma^\alpha
p_\alpha$. In 5D hyperbolic space the inner product of two vectors
can be null when the vectors are perpendicular but also when the two
vectors are null; since we have established that $p$ is a null
vector, Eq.\ (\ref{eq:nullcond}) can be satisfied either by
$\mathrm{d}x$ normal to $p$ or by $(\mathrm{d}x)^2 = 0$. In the
former case the condition describes a 3-volume called wavefront and
in the latter case it describes the wave motion. Notice that the
wavefronts are not surfaces but volumes, because we are working with
4-dimensional waves.

The condition describing wave motion can be expanded as
\begin{equation}
    \label{eq:wavemotion}
    -(\mathrm{d}x^0)^2 + \sum (\mathrm{d}x^i)^2 = 0.
\end{equation}
This is a purely scalar equation and can be manipulated as such,
which means we are allowed to rewrite it with any chosen terms in
the second member; some of those manipulations are particularly
significant. Suppose we decide to isolate $(\mathrm{d}x^4)^2$ in the
first member: $(\mathrm{d}x^4)^2 = (\mathrm{d}x^0)^2 - \sum
(\mathrm{d}x^m)^2$. We can then rename coordinate $x^4$ as $\tau$,
to get the interval squared of special relativity for space-like
displacements
\begin{equation}
    \mathrm{d}\tau^2 = (\mathrm{d}x^0)^2 - \sum
(\mathrm{d}x^m)^2.
\end{equation}
We have thus derived the space-like part of special relativity as a
consequence of monogeneity in 5D hyperbolic space and simultaneously
found a physical interpretation for coordinates $x^0$ and $x^4$ as
time and proper time, respectively.

A different manipulation of Eq.\ (\ref{eq:wavemotion}) has great
significance because it leads to the concept of \emph{4-dimensional
optics} (4DO).\cite{Almeida02:2, Almeida01} If we isolate
$(\mathrm{d}x^0)^2$ and replace $x^0$ by the letter $t$, we see that
time becomes the interval in Euclidean 4D space
\begin{equation}
    \mathrm{d}t^2 = \sum (\mathrm{d}x^i)^2.
\end{equation}
From this we conclude that the monogenic condition produces plane
waves whose wavefronts are 3D volumes but can be represented by
wavefront normals, just as it happens in standard optics with
electromagnetic waves.

Several readers may be worried with the fact that proper time is a
line integral and not a coordinate in special relativity; to this we
will argue that the manipulations we have done, collapsing 5D
spacetime into 4 dimensions through a null displacement condition
and then promoting one of the coordinates into interval, is exactly
equivalent to the process of defining a light cone in Minkowski
spacetime and then applying Fermat's principle to define an
Euclidean 3D metric on the light cone; we have just upgraded the
procedure by including one extra dimension.

\section{Quantum mechanics and the Dirac equation}

The Dirac equation can also be derived from the monogenic condition
but since it appears formulated in terms of matrices in all
textbooks we will have to rewrite Eq.\ (\ref{eq:monogenic}) also in
terms of matrices, so that it can then be further manipulated. This
is easily achieved if we assign our frame vectors to Dirac matrices
that square to the the identity matrix or minus the identity matrix
as appropriate; the following list of assignments can be used but
others would be equally effective\endnote{There are 16 possible $4
\times 4$ Dirac matrices,\cite{DiracMat} of which we must choose 5
such that $(\sigma_0)^2 = -I$, $(\sigma_i)^2 =I$ and $\sigma_\alpha
\sigma_\beta = -\sigma_\beta \sigma_\alpha$, for $\alpha \neq
\beta$.}
\begin{eqnarray}
\label{eq:diracmatrix}
   \sigma^0 &\equiv &  \left(\begin{array}{cccc} \mathrm{i} & 0 & 0 & 0 \\
    0 & -\mathrm{i} & 0 & 0 \\ 0 & 0 & \mathrm{i} & 0 \\ 0 & 0 & 0 &
    -\mathrm{i}
    \end{array}\right),~~
    \sigma^1 \equiv  \left(\begin{array}{cccc} 0 & 0 & 0 & 1 \\
    0 & 0 & -1 & 0 \\ 0 & -1 & 0 & 0 \\ 1 & 0 & 0 & 0
    \end{array}\right), \nonumber\\
    \sigma^2 &\equiv &   \left(\begin{array}{cccc} 0 & \mathrm{i} & 0 & 0 \\
    -\mathrm{i} & 0 & 0 & 0 \\ 0 & 0 & 0 & -\mathrm{i} \\ 0 & 0 & \mathrm{i} & 0
    \end{array}\right),~~
    \sigma^3 \equiv \left(\begin{array}{cccc} 0 & 1 & 0 & 0 \\
    1 & 0 & 0 & 0 \\ 0 & 0 & 0 & 1 \\ 0 & 0 & 1 & 0
    \end{array}\right), \nonumber \\
    \sigma^4 &\equiv &  \left(\begin{array}{cccc} 0 & 0 & 0 & -\mathrm{i} \\
    0 & 0 & \mathrm{i} & 0 \\ 0 & -\mathrm{i} & 0 & 0 \\ \mathrm{i} & 0 & 0 & 0
    \end{array}\right) .
\end{eqnarray}
There is no need to adopt different notations to refer to the frame
vectors or to their matrix counterparts because the context will
usually be sufficient to determine what is meant.

We can check that matrices $\sigma^\alpha$ form an orthonormal basis
of 5D space by defining the inner product of square matrices as
\begin{equation}
    A \dprod B = \frac{A B + B A}{2}.
\end{equation}
It will then be possible to verify that the inner product of any two
different $\sigma$-matrices is null, $(\sigma^0)^2 = -I$ and
$(\sigma^i)^2 = I$; these are the conditions defining an orthonormal
basis expressed in matrix form. \endnote{A more formal approach to
this subject would lead us to invoke the isomorphism between the
complex algebra of $4 \times 4$ matrices and Clifford algebra
$\mathcal{C}\ell_{4,1}$, the geometric algebra of 5D
spacetime.\cite{Lounesto01}}

It will now be convenient to expand the monogenic condition
(\ref{eq:monogenic}) as $(\sigma^\mu \partial_\mu + \sigma^4
\partial_4) \psi = 0$. If this is applied to the solution
(\ref{eq:psidef}) and the derivative with respect to $x^4$ is
evaluated we get
\begin{equation}
    (\sigma^\mu \partial_\mu + \sigma^4 \mathrm{i} p_4) \psi = 0.
\end{equation}
Let us now multiply both sides of the equation on the left by
$\sigma^4$ and note that matrix $\sigma^4 \sigma^0$ squares to the
identity while the 3 matrices $\sigma^4 \sigma^m$ square to minus
identity; we rename these products as $\gamma$-matrices in the form
$\gamma^\mu = \sigma^4 \sigma^\mu$. Rewriting the equation in this
form we get
\begin{equation}
    (\gamma^\mu \partial_\mu + i p_4) \psi = 0.
\end{equation}
The only thing this equation needs to be recognized as Dirac's is
the replacement of $p_4$ by the particle's mass $m$.

We turn now our attention to the amplitude $\psi_0$ in Eq.\
(\ref{eq:solution}) because we know that the Dirac equation accepts
solutions which are spinors and we want to find out their
equivalents in our formulation. Applying the monogenic condition to
Eq.\ (\ref{eq:solution}) we see that the following equation must be
verified
\begin{equation}
    \psi_0 (\sigma^\alpha p_\alpha) = 0.
\end{equation}
If the $\sigma$s are interpreted as matrices, remembering that $p$
is null, the only way the equation can be verified is by $\psi_0$
being some constant multiplied by the matrix in parenthesis, which
is a matrix representation of $p$. We can set the multiplying
constant to unity and $\psi_0$ becomes equal to $p$; the
wavefunction $\psi$ can then be interpreted as a Dirac spinor.

In order to separate \emph{left} and \emph{right} spinor components
we use a technique adapted from Ref.\ \onlinecite{Doran03}. We
choose an arbitrary $4 \times 4$ matrix which squares to identity,
for instance $\sigma_4$, with which we form the two idempotent
matrices $(I + \sigma_4)/2$ and $(I - \sigma_4)/2$. These matrices
are called idempotents because they reproduce themselves when
squared. These idempotents absorb any $\sigma_4$ factor; as can be
easily checked $(I + \sigma_4) \sigma_4 = (I + \sigma_4)$ and $(I -
\sigma_4) \sigma_4 = - (I - \sigma_4)$.

Obviously we can decompose the wavefunction $\psi$ as
\begin{equation}
    \psi =   \psi \frac{I + \sigma_4}{2} + \psi \frac{I -
    \sigma_4}{2} = \psi_+ + \psi_-.
\end{equation}
This apparently trivial decomposition produces some surprising
results due to the following relations
\begin{eqnarray}
    \mathrm{e}^{\mathrm{i} \theta} (I + \sigma_4) &=&
    (\cos \theta + \mathrm{i} \sin \theta) (I + \sigma_4) \nonumber \\
    &=& (I \cos \theta + \mathrm{i} \sigma_4 \sin \theta)
    (I + \sigma_4) \\
    &=& \mathrm{e}^{\mathrm{i}\sigma_4 \theta} (I + \sigma_4).
    \nonumber
\end{eqnarray}
and similarly
\begin{equation}
    \mathrm{e}^{\mathrm{i} \theta} (I - \sigma_4)=
    \mathrm{e}^{-\mathrm{i} \sigma_4 \theta} (I - \sigma_4).
\end{equation}
The different idempotents produce similar results and it has been
argued that they may be related to different elementary
particles.\cite{Almeida05:1}

\section{Conclusion}
The choice of an adequate geometry to write the fundamental
equations of physics is important in order to make the equations as
simple as possible. The perception one has of those equations is
also greatly dependent on the geometry and on the assignment between
coordinates and physical entities. This paper makes use of
5-dimensional spacetime and studies a special class of functions,
called monogenic functions, demonstrating that this is sufficient
for the derivation of equations in special relativity and quantum
mechanics and can thus be seen as a unifying principle between those
two areas of physics.

Monogenic functions in 5D spacetime produce consequences to 4D
Euclidean space and Minkowski spacetime simultaneously, which
provides two different points of view from which to perceive the
physical meaning of the solutions. For instance, solutions that can
be interpreted as Dirac particles in Minkowski spacetime are also 4D
"plane like" waves in Euclidean space; although the paper does not
explore this latter point of view, other works by the author have
shown that it provides a different and interesting perception of
quantum mechanics.

\begin{appendix}
\section{Non-dimensional units \label{units}}
The interpretation of $t$ and $\tau$ as time coordinates implies the
use of a scale parameter which is naturally chosen as the vacuum
speed of light $c$. We don't need to include this constant in our
equations because we can always recover time intervals, if needed,
introducing the speed of light at a later stage. We can even go a
step further and eliminate all units from our equations so that they
become pure number equations; in this way we will avoid cumbersome
constants whenever coordinates have to appear as arguments of
exponentials or trigonometric functions. We note that, at least for
the macroscopic world, physical units can all be reduced to four
fundamental ones; we can, for instance, choose length, time, mass
and electric charge as fundamental, as we could just as well have
chosen others. Measurements are then made by comparison with
standards; of course we need four standards, one for each
fundamental unit. But now note that there are four fundamental
constants: Planck constant $(\hbar)$, gravitational constant $(G)$,
speed of light in vacuum $(c)$ and proton electric charge $(e)$,
with which we can build four standards for the fundamental units.
\begin{table}[h]
\caption{\label{t:standards}Standards for non-dimensional units'
system}
\begin{center}
\begin{tabular}{c|c|c|c}
Length & Time & Mass & Charge \\
\hline & & & \\

$\displaystyle \sqrt{\frac{G \hbar}{c^3}} $ & $\displaystyle
\sqrt{\frac{G \hbar}{c^5}} $  & $\displaystyle \sqrt{\frac{ \hbar c
}{G}} $  & $e$
\end{tabular}
\end{center}
\end{table}
Table \ref{t:standards} lists the standards of this units' system,
frequently called Planck units, which the author prefers to
designate by non-dimensional units. In this system all the
fundamental constants, $\hbar$, $G$, $c$, $e$, become unity, a
particle's Compton frequency, defined by $\nu = mc^2/\hbar$, becomes
equal to the particle's mass and the frequent term ${GM}/({c^2 r})$
is simplified to ${M}/{r}$. We can, in fact, take all measures to be
non-dimensional, since the standards are defined with recourse to
universal constants; this will be our posture. Geometry and physics
become relations between pure numbers, vectors, bivectors, etc. and
the geometric concept of distance is needed only for graphical
representation.
\end{appendix}

\bibliography{Abrev,aberrations,assistentes}

\end{document}